\documentclass[showpacs,preprintnumbers,final,twocolumn]{revtex4}%
\usepackage{amssymb}
\usepackage{amsmath}
\usepackage{graphicx}
\usepackage{dcolumn}
\usepackage{bm}
\usepackage{amsfonts}
\usepackage{setspace}%
\providecommand{\U}[1]{\protect\rule{.1in}{.1in}}
\begin{document}
\title{Metastability in a nano-bridge based hysteretic DC-SQUID embedded in
superconducting microwave resonator}
\author{Eran Segev}
\email{segeve@tx.technion.ac.il}
\author{Oren Suchoi}
\author{Oleg Shtempluck}
\author{Fei Xue}
\altaffiliation{Current address: University of Basel.}

\author{Eyal Buks}
\affiliation{Department of electrical engineering, Technion, Haifa 32000, Israel}
\date{\today }

\begin{abstract}
We study the metastable response of a highly hysteretic DC-SQUID made of a
Niobium loop interrupted by two nano-bridges. We excite the SQUID with an
alternating current and with direct magnetic flux, and find different
stability zones forming diamond-like structures in the measured voltage across
the SQUID. When such a SQUID is embedded in a transmission line resonator
similar diamond structures are observed in the reflection pattern of the
resonator. We have calculated the DC-SQUID stability diagram in the plane of
the exciting control parameters, both analytically and numerically. In
addition, we have obtained numerical simulations of the SQUID equations of
motion, taking into account temperature variations and non-sinusoidal
current-phase relation of the nano-bridges. Good agreement is found between
experimental and theoretical results.

\end{abstract}

\pacs{85.25.Dq, 74.40.Gh, 74.25.Sv}
\maketitle





\section{Introduction}

In the last two decades Superconducting Quantum Interference Devices (SQUIDs)
have regained the interest of researchers worldwide, due to their use as
solid-state quantum bits. More recently, such SQUIDs were embedded in
superconducting Transmission Line Resonators (TLRs), in order to produce
circuit cavity quantum electrodynamics in the strong
\cite{Chiorescu2004,Johansson2006b,Houck2007,Majer2007a,Sillanpaa2007a,Astafiev2010}
and the dispersive
\cite{QubitResCoup_Lupacu06,qubitResReadout_Lee07,Lupascu2004} coupling
regimes \cite{Schoelkopf2008}. Other applications, in which SQUIDs are mainly
used as nonlinear classical elements, include Josephson bifurcation amplifiers
\cite{Mallet2009,Bergeal2010,Vijay2009,Astafiev2010a} and tunable resonators
\cite{Sandberg2008a,Castellanos-Beltran2007_083509,Laloy2008}. Tunable
resonators were also used to demonstrate parametric amplification and
squeezing
\cite{CASTELLANOS-BELTRAN2008,Castellanos-Beltran2009_944,Tholen2009_014019,Yamamoto2008_042510}%
, and might also be used to demonstrate the dynamical Casimir effect
\cite{Johansson2009,Wilson2010_2540v1}.

The injected power into a TLR is usually limited by the critical current of
the DC-SQUID embedded in the resonator. The upper bound of this current is
determined by the sum of the two critical currents of the Josephson junctions
(JJs) composing the DC-SQUID. While typical critical currents of DC-SQUIDs
range between few to tenth of microamperes, the applications that involve
resonance tuning and parametric amplification could benefit from larger
critical currents. In order to have larger critical currents one can fabricate
DC-SQUIDs using nano-bridges instead of JJs. Nano-bridges, which are merely
artificial weak-links having sub-micron size, were shown to have similar
Current-Phase Relationship (CPR) as JJs under certain conditions
\cite{Likharev1979_101,Troeman_024509,Hasselbach2002_4432}. These nano-bridge
JJs (NBJJs) are characterized by large critical current, on the order of
milliamperes \cite{Troeman2007,Hao2009_064011}. DC-SQUIDs with large critical
currents are often characterized by hysteretic response and metastable
dynamics
\cite{Tesche1977_301,Matsinger1978_91-96,Lefevre-Seguin1992_5507,Goldman1965_495,Palomaki2006_014520}%
. These characters are naturally made extreme in NBJJ based DC-SQUIDs. In
addition, NBJJs have very high plasma frequency \cite{Suchoi2010_174525}, on
the order of one Terahertz, which enables operation of the microwave TLR
without introducing interstate transitions in the embedded SQUID. Thus one
could employ a SQUID\ as a nonlinear lumped inductor, operating at microwave frequencies.

In this paper we experimentally and numerically study metastable response of
NBJJs based DC-SQUID, subjected to an alternating biasing current. We first
theoretically analyze stability zones of a highly hysteretic DC-SQUID in the
plane of the bias current and magnetic flux control parameters. Then we
directly measure the voltage across a SQUID in that plane. Comparison between
experimental results and between analytical and numerical theoretical
predictions yields good agreement. Moreover, we measure the reflection spectra
from several devices integrating a SQUID and a TLR and find qualitative
agreement between the theory and the experiments.

\section{Experimental Setup}%

\begin{figure}
[ptb]
\begin{center}
\includegraphics[
height=2.748in,
width=3.3441in
]%
{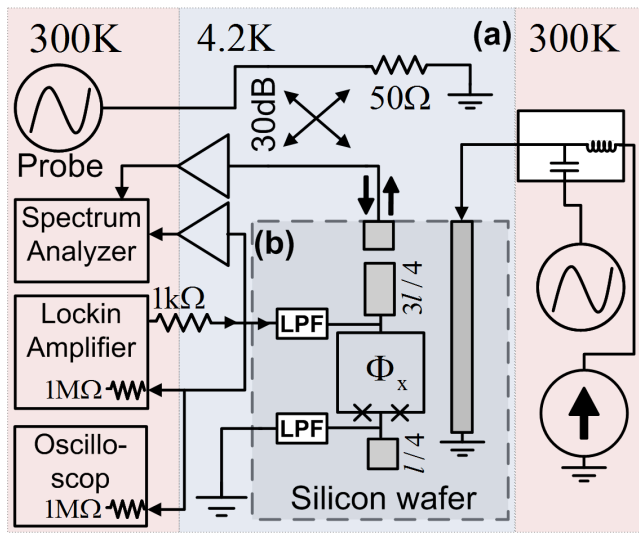}%
\caption{(Color online) $\mathrm{(a)}$ Measurement setup. $\mathrm{(b)}$
Schematic (unscaled) layout of our device. Red and blue background colors
indicate experimental setup having temperatures of $300\operatorname{K}$ and
$4.2\operatorname{K}$, respectively.}%
\label{ExpSetup}%
\end{center}
\end{figure}

A simplified circuit layout of our devices is illustrated in Fig.
\ref{ExpSetup}$\mathrm{(b)}$. We fabricate our devices on high resistivity
Silicon wafers, each covered by a thin layer of Silicon Nitride. Each device
is made of\ Niobium, having layer thickness of less than $100%
\operatorname{nm}%
$, and is composed of a stripline resonator having a DC-SQUID\ embedded in its
structure. The resonator is designed to operate in the gigahertz range, having
a length of about $l=19%
\operatorname{mm}%
$, which sets its first resonance mode around $2.5%
\operatorname{GHz}%
$. The DC-SQUID has two NBJJs, one NBJJ in each of its two arms (see Fig.
\ref{TheDevice}$\mathrm{(b)}$ and inset of Fig. \ref{squidPotential}). The
NBJJs are fabricated using FEI Strata $400$ focused ion beam system
\cite{Hao2007,Troeman2007} at an accelerating voltage of $30%
\operatorname{kV}%
$ and Gallium ions current of $1.5\mathrm{pA}$. The outer dimensions of the
bridges range from $100\times100%
\operatorname{nm}%
^{2}$ for large junctions to $60\times80%
\operatorname{nm}%
^{2}$ for relatively small junctions. The actual dimensions of the weak-links
are smaller because the bombarding Gallium ions penetrate into the Niobium
layer, and consequently suppress superconductivity over a depth estimated
between $30%
\operatorname{nm}%
$ and $50%
\operatorname{nm}%
$ \cite{Troeman2007,Datesman2005a,Tettamanzi2009_5302}. Despite the small
dimensions of our NBJJs, most of our SQUIDs have critical currents on the
order of milliamperes (see table \ref{Device Parameters}). A feed-line, weakly
coupled to the resonator, is employed for delivering input and output signals.
An on-chip transmission line passes near the DC-SQUID and is used to apply
magnetic flux through the DC-SQUID at frequencies ranging from DC to the
gigahertz. An on-chip filtered DC bias line is connected directly to the
DC-SQUID and is used for direct measurements of the SQUID. The Low-Pass
Filters (LPFs) are designed to minimize the degrading effect of these
connections on the quality factor of the resonator. Some measurements are
carried out while the device is fully immersed in liquid Helium, while others
are carried out in a dilution refrigerator where the device is in vacuum.
Further design considerations and fabrication details can be found elsewhere
\cite{Segev2009_152509,Suchoi2010_174525}.%

\begin{table}
\renewcommand{\arraystretch}{1.5}
\centering\begin{tabular} {c || c | c | c}
\hline\bfseries Parameter & \bfseries E19 & \bfseries E38 & \bfseries E42 \\
\hline\hline\bfseries SQUID Type & \bfseries RF & \bfseries DC & \bfseries
DC  \\
\hline\bfseries SQUID Area $[\operatorname{\mu m}^{2}]$ & \bfseries
$1936$ & \bfseries$870$ & \bfseries$1057$ \\
\hline\bfseries Nb Thickness $[\mathrm{nm}]$ & \bfseries$50$ & \bfseries
$100$ & \bfseries$60$ \\
\hline\bfseries Self-Inductance $[\mathrm{pH}]$ & \bfseries$127$ & \bfseries
$112$ & \bfseries$141$ \\
\hline\bfseries$I_{\mathrm{c}} [\mathrm{mA}_{\mathrm{RMS}}%
] $ & --- & \bfseries$2.29$ & \bfseries$1.74$ \\
\hline\bfseries$\beta_{\mathrm{L}}$ Calc & --- & \bfseries$1089$ & \bfseries
$106$ \\
\hline\bfseries$\beta_{\mathrm{L}}$ Fit & --- & \bfseries$722$ & \bfseries
$83$ \\
\hline\bfseries$\widetilde{\beta}_{\mathrm{L}}$ Fit & $1.5$ & \bfseries
--- & \bfseries$45 , 35$ \\
\hline\bfseries$\alpha$ Fit & --- & \bfseries$0.026$ & \bfseries$0.032$ \\
\hline\end{tabular}
\caption
{SQUIDs parameters. The self-inductance was numerically calculated using FastHenry computer program \cite
{Kamon94fasthenry}.
The parameter $\beta_{\mathrm{L}}%
$ Calc was evaluated analytically using the measured critical current.
The parameters $\beta_{\mathrm{L}}$ Fit, $\widetilde{\beta_{L}}%
$ Fit, and $\alpha$ Fit
were evaluated according to fittings of stability diagrams to measured data. }
\label{Device Parameters}
\end{table}%

The experimental results in this paper are obtained from three devices whose
parameters are summarized in table \ref{Device Parameters}. The experiments
are carried out using the setup depicted in Fig. \ref{ExpSetup}$\mathrm{(a)}$.
We report on two types of experiments. In the first one we obtain low
frequency current-voltage measurements of the SQUID using the DC bias line,
while the resonator does not play any role. We use a lock-in amplifier, which
applies alternating current through the SQUID, having excitation frequencies
on the order of kilohertz. We measure the voltage across the SQUID using the
lock-in amplifier and, in addition, we record the spectral density of the
voltage using a spectrum analyzer and its time domain dynamics using an
oscilloscope. In the second type of experiments we investigate the response of
an integrated SQUID-TLR device to a monochromatic incident probe tone that
drives one of the resonance modes of the TLR. The reflected power spectrum is
recorded by a spectrum analyzer. In such experiments the DC bias line is left
floating, and does not play any role in the measurement. In both types of
experiments we apply DC magnetic flux through the SQUID, and in experiments
with TLRs we also add modulated magnetic flux at gigahertz frequencies.

\subsection*{Numerical method}

Simulations of DC-SQUID circuit model (see Fig. \ref{TheDevice}$\mathrm{(a)}$)
are done by numerically integrating its Equations Of Motion (EOMs); Eqs.
(\ref{eom gamma_1}) and (\ref{eom gamma_2}). We introduce a sinusoidal
excitation current to the EOMs and calculate the phases of the two NBJJs
composing the DC-SQUID, the DC-SQUID\ voltage versus time, and the Fourier
transform of this voltage at the frequency of excitation. The experimental
measurement frequency was usually in the order of $1%
\operatorname{kHz}%
$, which is about nine orders of magnitude smaller than the SQUID plasma
frequency, and about six orders of magnitude smaller than the time-scale of
thermal processes in the SQUID \cite{Tarkhov2008_241112}. Therefore, in order
to make the simulations feasible in terms of computation time, we have made
two ease assumptions. The first assumption is that the excitation frequency
used in simulation can be increased as long as the DC-SQUID follows this
excitation adiabatically. Adiabatic approximation of NBJJs based DC-SQUID is
thoroughly analyzed in Ref. \cite{Suchoi2010_174525}, where it is shown that
the plasma frequency of a NBJJs based DC-SQUID is expected to be in the order
of $1%
\operatorname{THz}%
$. Thus in practice, the excitation frequency used in simulation is set
between $100%
\operatorname{MHz}%
$ and $1%
\operatorname{GHz}%
$. The second assumption is that thermal processes have a negligible influence
on the experimentally measured dynamics of the SQUID, as long as\ the
excitation frequency is kept low. Therefore, although we use high excitation
frequency in the simulations according to the first assumption, the
temperature of the SQUID is held at base temperature and does not evolve with
the SQUID\ dynamics. On the other hand, in simulations related to measurements
done with an integrated TLR-SQUID device, is which the measurement frequency
is high, thermal effects are taken into account by including thermal EOMs; Eq.
(\ref{EOM_Theta1}).

\section{Theory of Hysteretic DC-SQUID}%

\begin{figure}
[ptb]
\begin{center}
\includegraphics[
height=1.9701in,
width=3.3441in
]%
{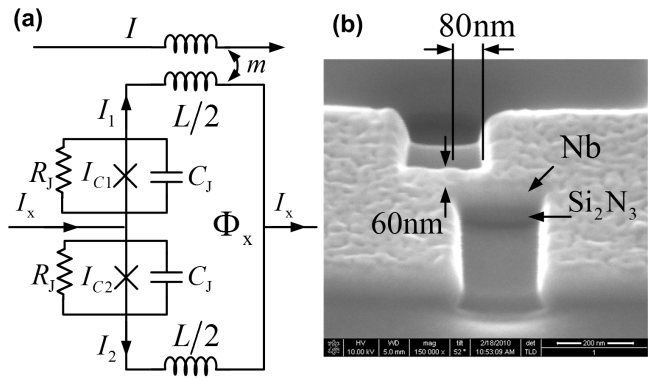}%
\caption{$\mathrm{(a)}$ Circuit model of a DC-SQUID. $\mathrm{(b)}$ Scanning
electron microscope (SEM) image of a nano-bridge, whose dimensions are
$80\times60\times50\operatorname{nm}^{3}$.}%
\label{TheDevice}%
\end{center}
\end{figure}

In this section, we develop a theory of hysteretic and metastable DC-SQUID,
taking into account the self-inductance and asymmetry of the DC-SQUID. Similar
models were also developed by others, but usually lacking comparison with
experimental data
\cite{Tesche1977_301,Goldman1965_495,Zahn1981_649,Tsang1975_4573,Palacios2006_021122,Matsinger1978_91,Ben-Jacob1983_6533}
or the emphasis put on other aspects of SQUID\ dynamics
\cite{Lefevre-Seguin1992_5507,Inchiosa1999_20,Wiesenfeld2000_R9232}. We begin
with a lossless model, in order to extract the SQUID\ stability diagram, and
add the dissipation and fluctuation terms to the EOMs on a later stage.

The circuit model is shown in Fig. \ref{TheDevice}$\mathrm{(a)}$ (a SEM image
can be seen in the inset of Fig. \ref{squidPotential}). It contains a
DC-SQUID\ having two NBJJs, one in each of its arms, with critical currents
$I_{\mathrm{c}1}$ and $I_{\mathrm{c}2}$, which in general may differ one from
another. Both NBJJs are assumed to have the same shunt resistance
$R_{\mathrm{J}}$ and capacitance $C_{\mathrm{J}}$ (which is considered
extremely small for NBJJs \cite{Ralph1996_10753}). The self-inductance of
DC-SQUID, $L$, is assumed to be equally divided between its two arms. Typical
inductance values of our SQUIDs are listed in table \ref{Device Parameters}.

The DC-SQUID is controlled by two external parameters. The first is bias
current $I_{\mathrm{x}}=I_{1}+I_{2}$, where $I_{1}$ and $I_{2}$ are the
currents flowing in the upper and lower arms respectively (Fig.
\ref{TheDevice}$\mathrm{(a)}$). The second is external magnetic flux
$\Phi_{\mathrm{x}}$ applied through the DC-SQUID. The total magnetic flux
threading the DC-SQUID loop is given by $\Phi=\Phi_{\mathrm{x}}+LI_{-},$where
$I_{-}=\left(  I_{1}-I_{2}\right)  /2\;$is circulating current in the loop.
Assuming sinusoidal CPR, the Josephson current $I_{\mathrm{J}k}$ in each
junction ($k=1,2$) is related to the critical current $I_{\mathrm{c}k}$ and to
the Josephson phase $\gamma_{k}$ by the Josephson equation $I_{\mathrm{J}%
k}=I_{\mathrm{c}k}\sin\gamma_{k}$. By employing the coordinate transformation
$\gamma_{+}=\left(  \gamma_{1}+\gamma_{2}\right)  /2$\ and $\gamma_{-}=\left(
\gamma_{1}-\gamma_{2}\right)  /2$, and the notation $I_{\mathrm{c}}=\left(
I_{\mathrm{c}1}+I_{\mathrm{c}2}\right)  $ and $I_{\mathrm{c}-}=\left(
I_{\mathrm{c}1}-I_{\mathrm{c}2}\right)  $, one finds that the potential
governing the dynamics of the DC-SQUID is given by
\cite{Lefevre-Seguin1992_5507}%

\begin{align}
\frac{u}{E_{0}}  &  =-\cos\gamma_{+}\cos\gamma_{-}+\alpha\sin\gamma_{+}%
\sin\gamma_{-}\nonumber\\
&  +\left(  \gamma_{-}+\frac{\pi\Phi_{\mathrm{x}}}{\Phi_{0}}\right)
^{2}/\beta_{\mathrm{L}}-\frac{I_{\mathrm{x}}}{I_{\mathrm{c}}}\gamma_{+}\;,
\label{u}%
\end{align}
where $\beta_{\mathrm{L}}=\pi LI_{\mathrm{c}}/\Phi_{0}$ is a dimensionless
parameter characterizing the DC-SQUID hysteresis, $\alpha=I_{\mathrm{c}%
-}/I_{\mathrm{c}}$ characterizes the DC-SQUID asymmetry, $E_{0}=\left(
\Phi_{0}I_{\mathrm{c}}\right)  /2\pi$ is the Josephson energy, and $\Phi_{0}$
is flux quantum.

\subsection{Stability Zones}

The extrema points of the DC-SQUID potential are found by solving%

\begin{align}
\frac{\partial u}{\partial\gamma_{+}}  &  =\sin\gamma_{+}\cos\gamma_{-}%
+\alpha\cos\gamma_{+}\sin\gamma_{-}-\frac{I_{\mathrm{x}}}{I_{\mathrm{c}}%
}=0\;,\label{u+=}\\
\frac{\partial u}{\partial\gamma_{-}}  &  =\cos\gamma_{+}\sin\gamma_{-}%
+\alpha\sin\gamma_{+}\cos\gamma_{-}+\frac{2\gamma_{-}}{\beta_{\mathrm{L}}%
}-\frac{2\pi\Phi_{\mathrm{x}}}{\Phi_{0}\beta_{\mathrm{L}}}\;\nonumber\\
&  =0. \label{u-=}%
\end{align}
In general, Eqs. (\ref{u+=}) and (\ref{u-=}) have periodic solutions, where
the solutions differ one from another by $2\pi m_{+}$ in $\gamma_{+}$, $2\pi
m_{-}$ in\ $\gamma_{-}$, and by $2\Phi_{0}m_{-}$ in $\Phi_{\mathrm{x}}$, where
$m_{+}$ and $m_{-}$ are integers.

The Jacobian of the potential $u$ is given by
\begin{equation}
J=\left(
\begin{array}
[c]{cc}%
\frac{\partial^{2}u}{\partial\gamma_{+}^{2}} & \frac{\partial^{2}u}%
{\partial\gamma_{+}\partial\gamma_{-}}\\
\frac{\partial^{2}u}{\partial\gamma_{-}\partial\gamma_{+}} & \frac
{\partial^{2}u}{\partial\gamma_{-}^{2}}%
\end{array}
\right)  \;.
\end{equation}
For local minima points of the potential, both eigenvalues of $J$ are
positive. Thus, we find boundaries of stability regions of these minima points
in the plane of the JJ phases, $\gamma_{+}$ and $\gamma_{-}$, by demanding
that
\begin{equation}
\det J=0\text{, }\operatorname{tr}J>0. \label{LSZbounderyCond}%
\end{equation}
Furthermore, the matrix $J$ is independent of both control parameters
$I_{\mathrm{x}}$ and $\Phi_{\mathrm{x}}$, thus also these boundaries are
independent of $I_{\mathrm{x}}$ and $\Phi_{\mathrm{x}}$. Finding the stability
thresholds in the plane of $I_{\mathrm{x}}$ and $\Phi_{\mathrm{x}}$ is done by
substituting the solutions of Eq. (\ref{LSZbounderyCond}) in Eqs. (\ref{u+=})
and (\ref{u-=}).%
\begin{figure}
[ptb]
\begin{center}
\includegraphics[
height=2.7023in,
width=3.3441in
]%
{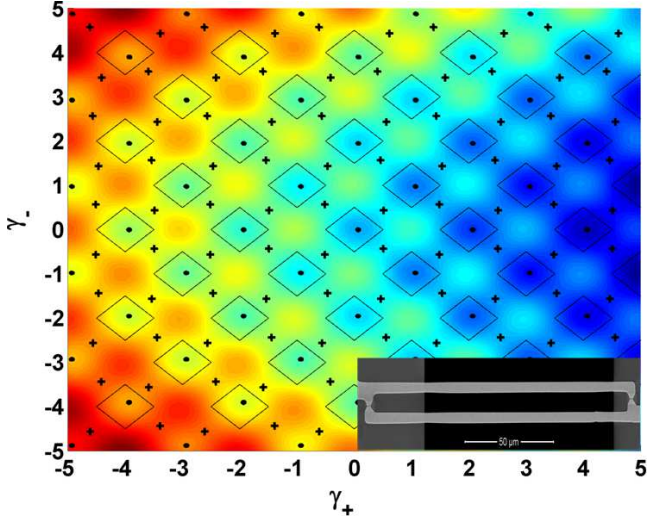}%
\caption{(Color online) Potential diagram of a DC SQUID, $u\left(  \gamma
_{+},\gamma_{-}\right)  \,$, drawn using $I_{\mathrm{x}}=0.1I_{\mathrm{c}}$,
and $\Phi_{\mathrm{x}}=0$. Local minima points are labeled by black dots and
saddle points are labeled by plus signs. The inset shows a SEM image of a
DC-SQUID.}%
\label{squidPotential}%
\end{center}
\end{figure}
\qquad

Figure \ref{squidPotential} plots the potential diagram of a DC-SQUID,
calculated for E42 using $\beta_{\mathrm{L}}=83$ and $\alpha=0.032$. The
solution of Eq. (\ref{LSZbounderyCond}) produces the black closed curves that
enclose the DC-SQUID local minima points in the plane of $\gamma_{+}$ and
$\gamma_{-}$. The local minima points are labeled by black dots and saddle
points are labeled by plus signs. A local minimum point losses its stability
when increasing either $\left\vert I_{\mathrm{x}}\right\vert $ or $\left\vert
\Phi_{\mathrm{x}}\right\vert $ to a point where it merges with one of the
saddle points close to it.

\subsubsection*{The limits of small and large $\beta_{\mathrm{L}}$}

In general, solving Eqs. (\ref{u+=}) and (\ref{u-=}) for a given set of
$\gamma_{+}$ and $\gamma_{-}$ can only be done numerically. Analytical
solution can be derived only for the extreme limits of $\beta_{\mathrm{L}}%
\ll1$ and $\beta_{\mathrm{L}}\gg1$. In the former limit, which is not the
focus of this paper, the derivation leads to the well known formula for the
SQUID critical current \cite{Clarke2004}%
\begin{equation}
\sqrt{1-\left(  1-\alpha^{2}\right)  \sin^{2}\frac{\pi\Phi_{\mathrm{x}}}%
{\Phi_{0}}}=\frac{I_{\mathrm{x}}}{I_{\mathrm{c}}}.
\end{equation}
In the opposite limit of $\beta_{\mathrm{L}}\gg1$, the condition of $\det J=0$
implies that $\cos\gamma_{1}\cos\gamma_{2}=0$, and the condition of
$\operatorname{tr}J>0$ implies that $\left(  1+\alpha\right)  \cos\gamma
_{1}+\left(  1-\alpha\right)  \cos\gamma_{2}>0$. Consider first the solution
for the minimum point near $\left(  \gamma_{+},\gamma_{-}\right)  =\left(
0,0\right)  $. Other solutions can be obtained from the periodic properties of
Eqs. (\ref{u+=}) and (\ref{u-=}). For this solution, Eq. \ref{LSZbounderyCond}
is satisfied along a square that is formed by the lines connecting the four
vertexes $\gamma_{1}=\pm\pi/2$ and $\gamma_{2}=\pm\pi/2$. Substituting these
vertexes into Eqs. (\ref{u+=}) and (\ref{u-=}) yields a bounding contour which
has a rectangle shape with vertexes at $\left(  I_{\mathrm{x}}/I_{\mathrm{c}%
},~2\pi\Phi_{\mathrm{x}}/\Phi_{0}\beta_{\mathrm{L}}\right)  =\left(
1,\alpha\right)  $, $\left(  \alpha,1\right)  $, $\left(  -1,-\alpha\right)  $
and $\left(  -\alpha,-1\right)  $ in the plane of the control parameters
$I_{\mathrm{x}}$ and $\Phi_{\mathrm{x}}$ (see Fig. \ref{stabZones}%
$\mathrm{(a)}$). This rectangle crosses the axis $I_{\mathrm{x}}=0$ at the
points $\Phi_{\mathrm{x}}=\pm\Phi_{0}\beta_{\mathrm{L}}\left(  1-\alpha
\right)  /2\pi$ and the axis $\Phi_{\mathrm{x}}=0$ at the points
$I_{\mathrm{x}}=\pm I_{\mathrm{c}}\left(  1-\alpha\right)  $.

\subsection{Equations of motion}

Applying Kirchhoff's laws on the DC-SQUID circuit model, substituting
Josephson's current-phase and voltage-phase equations, and taking into account
the fluctuation dissipation theory, yields the following EOMs for the
SQUID\ phases $\gamma_{1}$ and $\gamma_{2}$ \cite{Clarke2004}
\begin{gather}
\ddot{\gamma}_{1}+\beta_{\mathrm{D}}\dot{\gamma}_{1}+\left(  1+\alpha
_{0}\right)  y\left(  \Theta_{1}\right)  \sin\gamma_{1}\nonumber\\
+\frac{1}{\beta_{\mathrm{L0}}}\left(  \gamma_{1}-\gamma_{2}+2\pi
\Phi_{\mathrm{x}}/\Phi_{0}\right)  =I_{\mathrm{x}}/I_{\mathrm{c}%
0}+g_{\mathrm{n}1}, \label{eom gamma_1}%
\end{gather}%
\begin{gather}
\ddot{\gamma}_{2}+\beta_{\mathrm{D}}\dot{\gamma}_{2}+\left(  1-\alpha
_{0}\right)  y\left(  \Theta_{2}\right)  \sin\gamma_{2}\nonumber\\
-\frac{1}{\beta_{\mathrm{L0}}}\left(  \gamma_{1}-\gamma_{2}+2\pi
\Phi_{\mathrm{x}}/\Phi_{0}\right)  =I_{\mathrm{x}}/I_{\mathrm{c}%
0}+g_{\mathrm{n}2}, \label{eom gamma_2}%
\end{gather}
where the overdot denotes a derivative with respect to a normalized time
parameter $\tau=\omega_{\mathrm{pl}}t$, where $\omega_{\mathrm{pl}}$ is the
SQUID plasma frequency, and $\beta_{\mathrm{D}}=1/\left(  R_{\mathrm{J}%
}C_{\mathrm{J}}\omega_{\mathrm{pl}}\right)  $ is the damping coefficient. In
general, the NBJJ critical current, $I_{\mathrm{c}k}$ $(k=1,2)$, and thus
$I_{\mathrm{c}}$ and $I_{\mathrm{c-}}$, are temperature dependant, as we
discuss later. Thus we employ the notation $I_{\mathrm{c0}k}$, $I_{\mathrm{c0}%
}$, and $I_{\mathrm{c0-}}$ for the corresponding critical currents at a base
temperature $T_{0}$, which is the temperature of the coolant. In addition, we
employ the notation $\beta_{\mathrm{L0}}=\pi LI_{\mathrm{c0}}/\Phi_{0}$ and
$\alpha_{0}=\left(  I_{\mathrm{c0}1}+I_{\mathrm{c0}2}\right)  /\left(
I_{\mathrm{c0}1}-I_{\mathrm{c0}2}\right)  $. The term $y\left(  \Theta
_{k}\right)  $, where $\Theta_{k}=T_{k}/T_{\mathrm{c}}$ is the normalized
temperature of the $k^{\mathrm{th}}$ NBJJ, expresses the dependence of the
NBJJ\ critical current on its temperature, and is equal to unity as long as
the temperature of the NBJJs are held at base temperature. The factor
$g_{\mathrm{n}k}$ is a noise term, whose spectral density for the case where
$h\nu/k_{\mathrm{B}}T\ll1$ is given by $S_{I_{\mathrm{n}}}\left(  \nu\right)
=4k_{\mathrm{B}}T/R_{\mathrm{J}}$, with $k_{\mathrm{B}}$ being the Boltzmann
constant. In what follows we neglect this noise term in the numerical simulations.

To evaluate the voltage across the DC-SQUID, denoted as $V_{\mathrm{SQD}}$, we
assume that the loop inductance is equally divided between its two arms, and
get
\begin{equation}
V_{\mathrm{SQD}}=\frac{1}{2}\left[  \frac{\Phi_{0}}{2\pi}\left(
\frac{\mathrm{d}\gamma_{1}}{\mathrm{d}t}+\frac{\mathrm{d}\gamma_{2}%
}{\mathrm{d}t}\right)  +\frac{L}{2}\frac{\mathrm{d}I_{\mathrm{x}}}%
{\mathrm{d}t}\right]  . \label{Vs}%
\end{equation}

It is known that NBJJs may have complex CPR
\cite{Beenakker1991_3056,Golubov2004,Troeman_024509}, which deviates from the
normal sinusoidal CPR of a regular JJ. According to a theory, brought in the
appendix for completeness, such deviation would modify the $\sin\gamma_{i}$
term in the SQUID EOMs, Eqs. (\ref{eom gamma_1}) and (\ref{eom gamma_2}),
which become Eqs. (\ref{EOM_CPR1}) and (\ref{EOM_CPR2}). We have made some
simulations in which moderate changes in the CPR of our SQUIDs is assumed, and
found no significant difference between these results and results that neglect
this deviation. In addition, the physical dimensions of our NBJJs are
relatively small, and the measured values of $\beta_{\mathrm{L}}$ of our
SQUIDs are relatively high, thus following the explanations detailed in
appendix $\mathrm{A1}$ of Ref. \cite{Suchoi2010_174525}, the effect of
non-sinusoidal CPR in our NBJJs is expected to be negligibly small. Therefore,
our conclusion is that our SQUIDs can be modeled by normal sinusoidal CPR.

\subsection{Stability Diagram}%

\begin{figure}
[ptb]
\begin{center}
\includegraphics[
trim=0.000000in 0.000000in 0.000000in -0.312192in,
height=2.5081in,
width=3.3441in
]%
{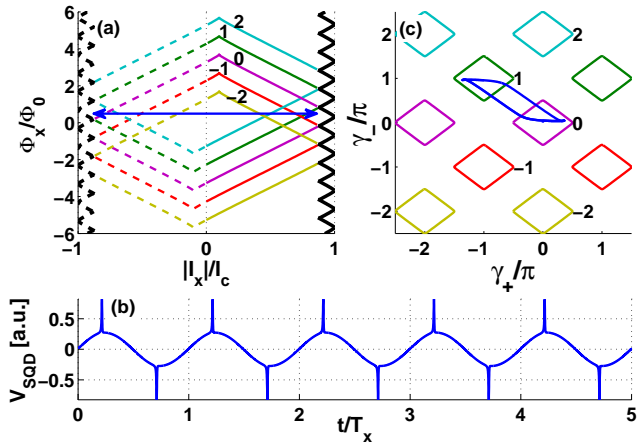}%
\caption{(Color online) $\mathrm{(a)}$ Stability diagram in the plane of the
control parameters $I_{\mathrm{x}}$ and $\Phi_{\mathrm{x}}$, drawn for
$\beta_{L}=20$ and $\alpha=0.1$. The curves are drawn using solid line for
$I_{\mathrm{x}}>0$ and dashed line for $I_{\mathrm{x}}<0$. The number close to
each curve labels the number of flux quanta trapped in the DC-SQUID in the
corresponding LSZ. Corresponding numbers are also indicated in panel
$\mathrm{(c)}$. The bold black curve marks the threshold between static and
oscillatory zones. The double-headed blue arrow is drawn between the points
$I_{\mathrm{x}}=\pm0.9I_{\mathrm{c}}$, $\Phi_{\mathrm{x}}=0.52\Phi_{0}$, and
marks the control parameter range used in the numerical simulation shown in
panels $\mathrm{(b)}$ and $\mathrm{(c)}$. These panels show SQUID voltage in
the time domain $\mathrm{(b)}$ and the NBJJs\ phases in the phase space
$\mathrm{(c)}$. The time scale is normalized by the period of excitation
$T_{\mathrm{x}}=\omega_{\mathrm{x}}/2\pi$.}%
\label{stabZones}%
\end{center}
\end{figure}

The stability diagrams in the plane of the control parameters $I_{\mathrm{x}}$
and $\Phi_{\mathrm{x}}$, and in the plane of the DC-SQUID\ phases $\gamma_{+}$
and $\gamma_{-}$, are plotted in Fig. \ref{stabZones}$\mathrm{(a)}$ and
\ref{stabZones}$\mathrm{(c)}$, respectively. The curves were first calculated
in the plane of $\gamma_{+}$ and $\gamma_{-}$ by numerically solving Eq.
(\ref{LSZbounderyCond}), with the parameters $\beta_{\mathrm{L}}=20$ and
$\alpha=0.1$. The above solutions were then substituted into Eqs. (\ref{u+=})
and (\ref{u-=}) to produce the closed contours of panel $\mathrm{(a).}$ Each
closed contour bounds a Local Stability Zone (LSZ) corresponding to a
different integer number of flux quanta trapped in the DC-SQUID. Corresponding
LSZs in Fig. \ref{stabZones} $\mathrm{(a)}$ and \ref{stabZones}$\mathrm{(c)}$
are labeled by the same number, indicating the number of trapped flux quanta.

The stability diagram can be separated into two global zones. The first is
called static zone \cite{Inchiosa1999_20,Wiesenfeld2000_R9232}, where the
system has one or more LSZs depending of the value of the hysteresis parameter
$\beta_{\mathrm{L}}$. The static zone is bounded in the horizontal axis by a
threshold current called the oscillatory threshold, given by $I_{\mathrm{th}%
}\left(  \Phi_{\mathrm{x}}\right)  =I_{\mathrm{c}}-I_{-}\left(  \Phi
_{\mathrm{x}}\right)  $. This threshold is periodic on the external flux
$\Phi_{\mathrm{x}}$, having a maximum value equal to the critical current of
the DC-SQUID. When the DC-SQUID\ is biased to the static zone, it is always
found in a LSZ, though transitions between LSZs may be forced by the control
parameters. The second zone, called oscillatory zone
\cite{Inchiosa1999_20,Wiesenfeld2000_R9232} (also known as dissipative or
free-running zone), spreads over two unbounded areas for which the excitation
current is larger (in absolute value) than the oscillatory threshold. In this
zone the DC-SQUID has no stable state; it oscillates at very high frequencies
and dissipates energy. In what follows we focus our study on the static zone.
Further study of the dynamics in the region of spontaneous oscillations can be
found in Ref. \cite{Wiesenfeld2000_R9232}.

\subsection{SQUID Dynamics}

The EOMs, Eqs. (\ref{eom gamma_1}) and (\ref{eom gamma_2}), were numerically
integrated using the parameters which used to draw the stability diagram of
Fig. \ref{stabZones}, and using the control parameters $\Phi_{\mathrm{x}%
}=0.52\Phi_{0}$ and $I_{\mathrm{x}}/I_{\mathrm{c}}=0.9$. The range of the
excitation is marked by a double-headed, blue arrow in the stability diagram
of Fig. \ref{stabZones}$\mathrm{(a)}$. Note that the left arrow head crosses
the threshold separating LSZ\_0 and LSZ\_1 (LSZs corresponding to $0$ or $1$
trapped flux quanta, respectively); whereas the right headed arrow crosses the
threshold separating LSZ\_1 back into LSZ\_0. The results of a simulation in
which a DC-SQUID is periodically excited along this path are shown in Fig.
\ref{stabZones}. Panel $\mathrm{(b)}$ shows the DC-SQUID voltage in the time
domain. Each excitation cycle contains two spikes, which occur close to
extrema points of the excitation amplitude. Panel $\mathrm{(c)}$ shows the
simulation in the phase plane of $\gamma_{+}$ and $\gamma_{-}$. The simulation
shows that the system periodically switches between LSZ\_0 and LSZ\_1, and
that the dynamics does not involve any additional LSZs. Thus a positive spike
in the time-domain response corresponds to a transition from LSZ\_0 to LSZ\_1,
and a negative spike corresponds to an opposite transition. When the
excitation is monotonically increased the system mostly lingers in LSZ\_0 and
when it is decreased it mostly lingers in LSZ\_1. Note that a serial
resistor\ $R=0.5%
\operatorname{\Omega }%
$ was post-simulation added to the voltage results in order to emphasize the
excitation cycle. Such a serial resistance also unavoidably exists in the
wiring of our experimental setup. Note also that the duration of each spike,
which is related to the relaxation time of the SQUID, is negligible compared
to the period of excitation, $T_{\mathrm{x}}=\omega_{\mathrm{x}}/2\pi$. Thus,
the assumption that the response of the DC-SQUID to the excitation is
adiabatic is reasonable.

\subsection{Periodic dissipative static zone}%

\begin{figure}
[ptb]
\begin{center}
\includegraphics[
height=2.631in,
width=3.3441in
]%
{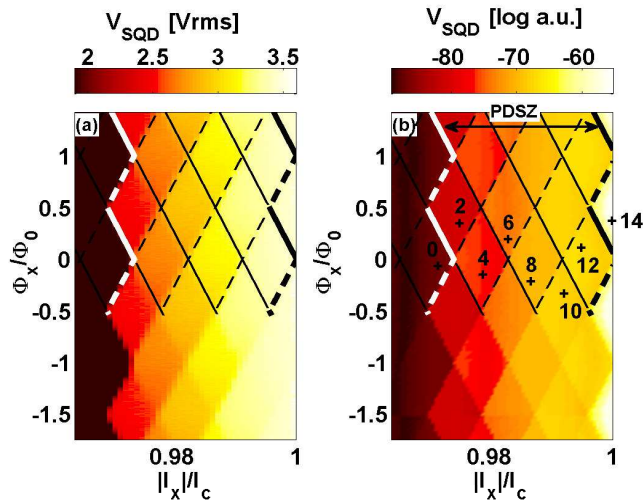}%
\caption{(Color online) Experimental $\mathrm{(a)}$ and numerical calculations
$\mathrm{(b)}$ of direct voltage measurements and calculations of E38, in the
parameter plane of $I_{\mathrm{x}}$ and $\Phi_{\mathrm{x}}$. Black contours
mark the corresponding stability diagram, drawn using solid line for
$I_{\mathrm{x}}>0$ and dashed line for $I_{\mathrm{x}}<0$. The diagram is
applied only on half of the colormap in order to leave some data uncovered.
The white bold contour marks the PDSZ threshold. Black bold contour marks the
oscillatory threshold. Each plus sign marks a set of control parameters, and
the corresponding number counts the LSZs boundaries that the SQUID
periodically crosses if excited by that set of parameters.}%
\label{vvsflux_simexp_3d}%
\end{center}
\end{figure}

Figure \ref{vvsflux_simexp_3d} shows a quantitative comparison between
experimental results measured with E38 using lock-in amplifier (panel
$\mathrm{(a)}$) and simulation results calculated using the corresponding
parameters (panel $\mathrm{(b)}$). Both panels show colormaps of the voltage
across the DC-SQUID as a function of control parameters, $I_{\mathrm{x}}$, and
$\Phi_{\mathrm{x}}$. The black contours in both graphs mark the corresponding
stability diagram. The negative part of the stability contours is folded onto
the positive part and is drawn by dashed lines. Thus, solid lines represent
stability thresholds for positive excitation currents, and dashed lines
represent stability thresholds for negative excitation currents. The bold
black contour is the threshold between the static zone and the oscillatory one.

The folding of the stability contours divides the static zone of the stability
diagram into regions having diamond shapes in the plane of $I_{\mathrm{x}}$
and $\Phi_{\mathrm{x}}$. Each diamond region bounds a range of parameters for
which the system crosses the same number of stability thresholds during an
excitation cycle. For example, when excited with parameters bounded by the
diamond marked by $2$, the DC-SQUID would cross a single threshold during the
positive duration of the excitation cycle and another one during the negative
duration, thus returning to the original LSZ, similar to the case discussed in
Fig. \ref{stabZones}$\mathrm{(c)}$. Each crossing of a threshold line, either
by positive or negative currents, triggers a spike, which in turn contributes
additively to the measured voltage across the DC-SQUID. Thus, each diamond
bounds a range of parameters for which a similar voltage is measured. The
diamonds follow the periodicity properties dictated for the control parameters
by Eqs. (\ref{u+=}) and (\ref{u-=}).

We define an additional threshold, marked by bold white contour, which
separates the static zone into two sections. In the first, which applies for
periodic excitation currents smaller (in absolute value) than this threshold,
the DC-SQUID is captured in a single LSZ after finite number of excitation
cycles, and no spikes in the voltage appear afterwards. We call this section
Periodic Non-Dissipative Static Zone (PNDSZ). In the second section, which
spreads for excitation currents larger than the white threshold but smaller
than the oscillatory threshold, the SQUID periodically jumps between LSZs and
dissipates energy. Thus, we call this section Periodic Dissipative Static Zone
(PDSZ), and call the white threshold itself the PDSZ threshold.%

\begin{figure}
[ptb]
\begin{center}
\includegraphics[
height=2.6127in,
width=3.3441in
]%
{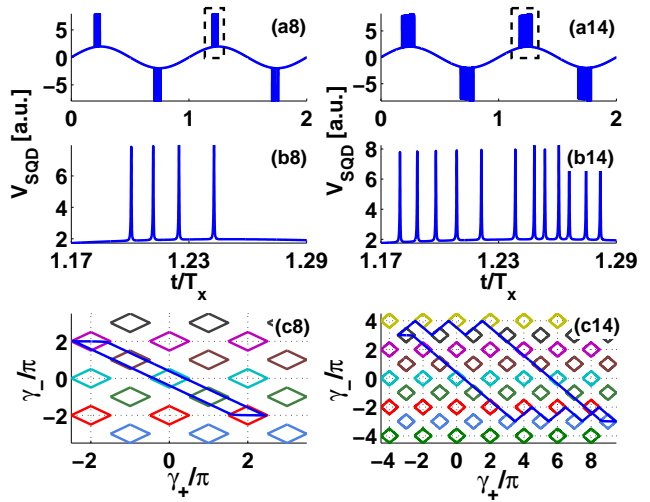}%
\caption{(Color online) Simulation results for E38. Panels $\mathrm{(a}%
i\mathrm{)}$, where $i=8,14$ corresponding to the marked points in Fig.
\ref{vvsflux_simexp_3d}, plot the DC-SQUID\ voltage in the time domain. panels
$\mathrm{(b}i\mathrm{)}$ magnify the corresponding marked squares in panels
$\mathrm{(a}i\mathrm{)}$. Panels $\mathrm{(c}i\mathrm{)}$ show the simulation
results and the corresponding stability zones in the phase space of
$\gamma_{+}$ and $\gamma_{-}$.}%
\label{TimeDomainSimEmb38}%
\end{center}
\end{figure}
\begin{figure*}
\centering
\[%
\begin{array}
[c]{cc}%
\text{%
{\includegraphics[
height=2.7298in,
width=3.3441in
]%
{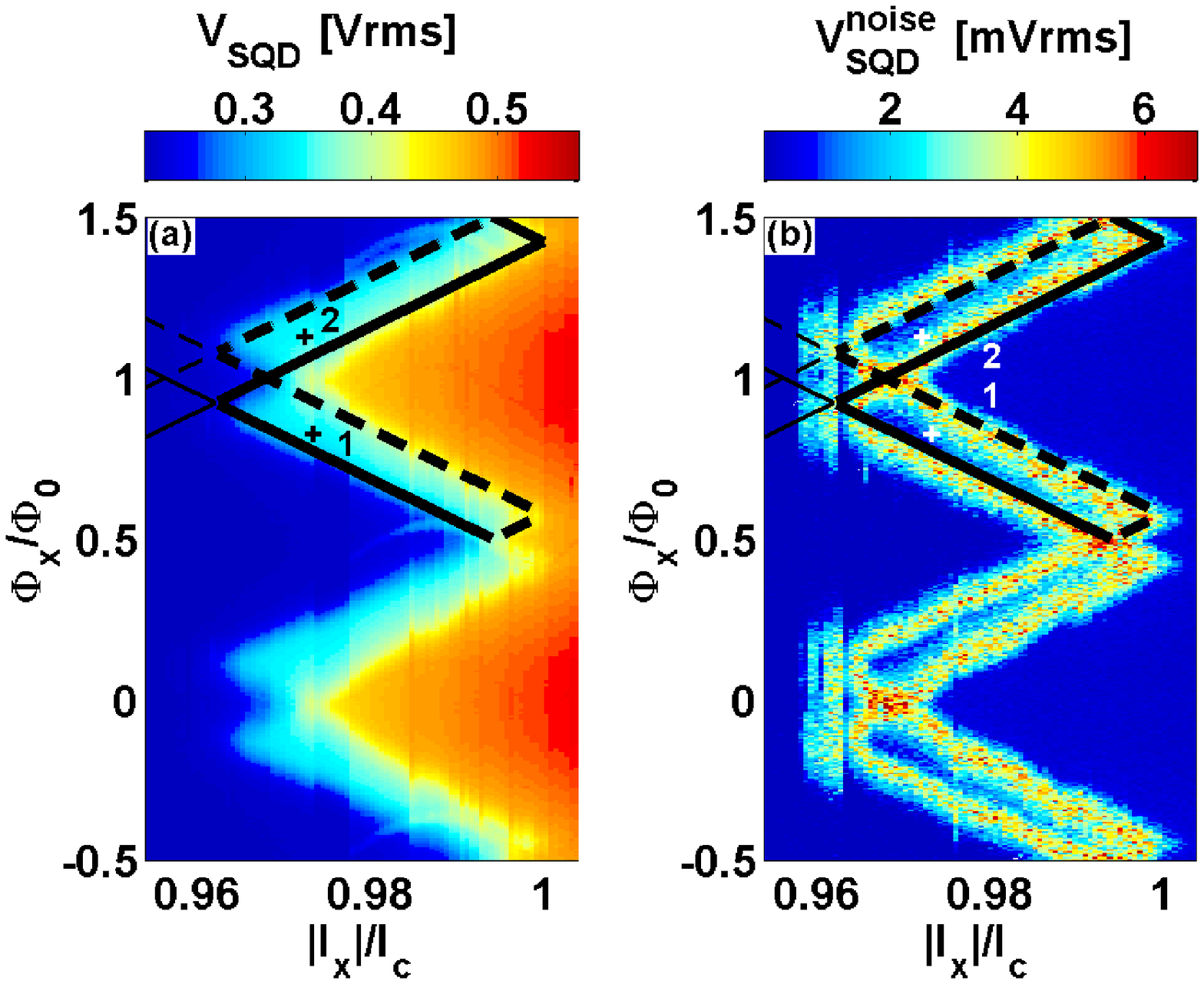}%
}
} & \text{%
{\includegraphics[
height=2.6874in,
width=3.3441in
]%
{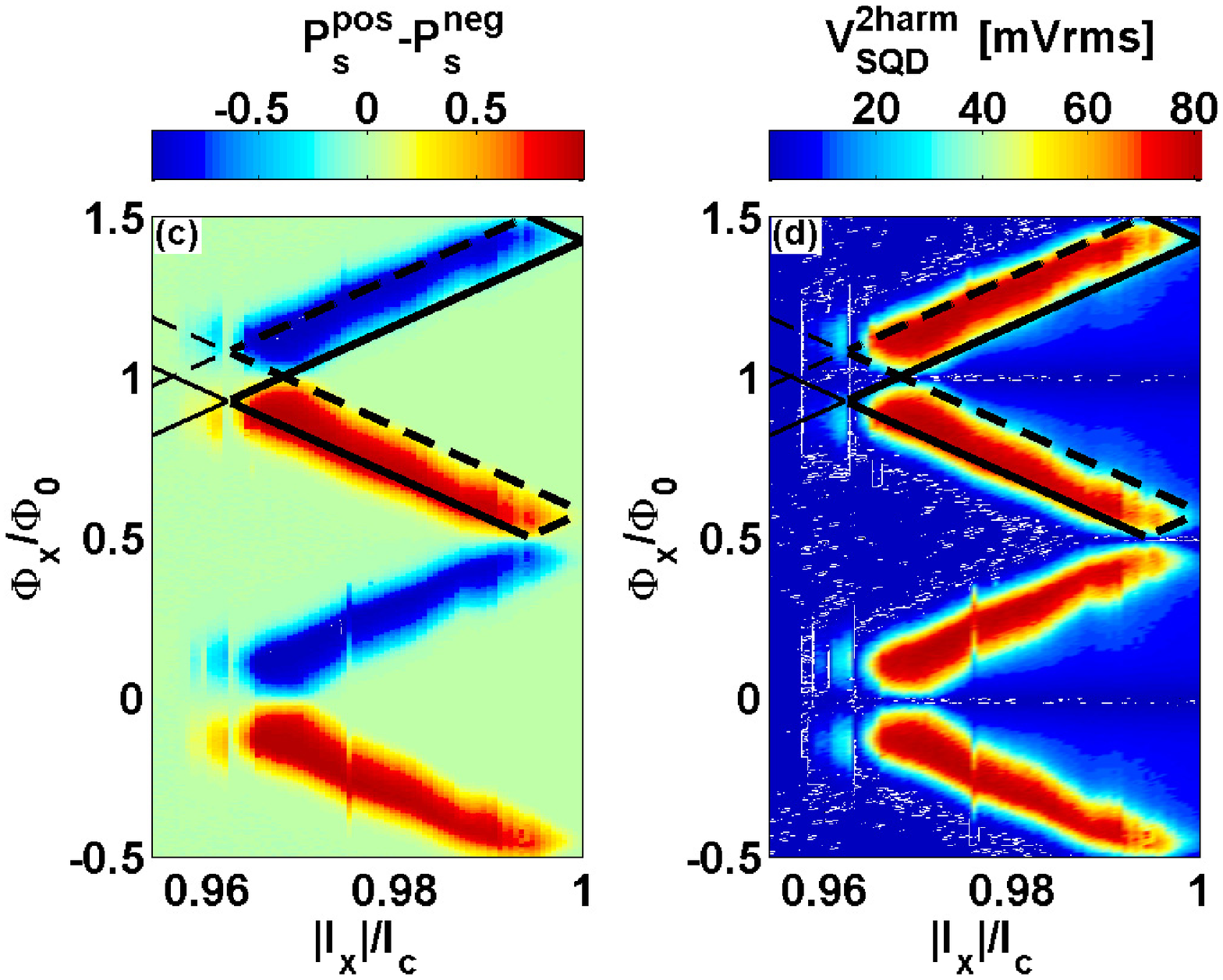}%
}
}%
\end{array}
\]
%

\caption{(Color online) Experimental measurements obtained from E42.
Panels $\mathrm{(a)}$ and  $\mathrm{(d)}%
$ draw the first and second SQUID voltage harmonics respectively.
Panel  $\mathrm{(b)}$ draws the voltage noise level at frequency of 1kHz.
Panel  $\mathrm{(c)}%
$ draws a statistical analysis of time-domain voltage traces,
which calculates the difference between probabilities of measuring positive and negative voltage spikes respectively.}%
\label{EMb42aExp}%
\end{figure*}%

Figure \ref{TimeDomainSimEmb38} shows results of two simulations, calculated
using the control parameters marked by points $8$ and $14$ in Fig.
\ref{vvsflux_simexp_3d}. Panels $\mathrm{(a}i\mathrm{)}$ where $i=8,14$ show
the DC-SQUID voltage as a function of time, calculated during two excitation
cycles. Each cycle has a bunch of spikes during its positive duration and
another bunch during the negative one. Panels $\mathrm{(b}i\mathrm{)}$ magnify
the corresponding bunch of spikes marked by dashed squares in panels
$\mathrm{(a}i\mathrm{)}$. In panel $\mathrm{(b8)}$ one counts four spikes
corresponding to four crossings of stability thresholds. Panel $\mathrm{(c8)}$
shows the phase space dynamics in the plane of $\gamma_{-}$ and $\gamma_{+}$.
The system periodically cycles between five LSZs, where most of the time it
lingers either in the upper left or in the lower right LSZs. The transition
between these LSZs is forced by the driving sinusoidal bias current, which
induces four jumps, corresponding to the four spikes in panel $\mathrm{(b8)}$.
Panels $\mathrm{(b14)}$ and $\mathrm{(c14)}$ show a richer dynamics, which
emerges from the fact that point $\mathrm{+14}$ is located inside the
oscillatory zone. The dynamics includes both forced transitions between LSZs
and spontaneous oscillations related to the oscillatory zone. These two kinds
of transitions are clearly seen in Fig. \ref{vvsflux_simexp_3d}, panel
$\mathrm{(c14)}$, which plots the dynamics of $\gamma_{+}$ and $\gamma_{-}$ in
the phase space. The first kind, in which $\gamma_{-}$ and $\gamma_{+}$ change
monotonically, corresponds to forced transitions between LSZs. The second
kind, in which the system cranks between LSZs, corresponds to the spontaneous
oscillatory dynamics. The spontaneous oscillations last as long as the
temporal driving bias current is greater than the oscillatory threshold. The
distinction between the forced transitions and the spontaneous oscillations
can also be noticed in the time domain, shown in Fig. \ref{vvsflux_simexp_3d},
panel $\mathrm{(b14)}$, where the first six spikes are similar and distinct,
whereas the rest of the spikes emerge in pairs, in which one spike is slightly
stronger than the other.

\subsection{Hybrid Oscillatory Zones}

Figure \ref{EMb42aExp} shows low-frequency experimental measurements of E42
DC-SQUID, which has a moderated self-inductance parameter $\beta_{\mathrm{L}%
}=83$ (compared to the one of E38). Similar to the experiment with E38, we
excite the SQUID by a sinusoidal current having frequency of $\omega
_{\mathrm{x}}/2\pi=2.5%
\operatorname{kHz}%
$, and measure the voltage across the DC-SQUID as a function of excitation
current $I_{\mathrm{x}}$ and external magnetic flux $\Phi_{\mathrm{x}}$. In
addition, we measure the spectral density of the voltage using a spectrum
analyzer and its response in the time domain using an oscilloscope. Figure
\ref{EMb42aExp}$\mathrm{(a)}$ shows this voltage and has the corresponding
folded stability diagram drawn on it. The colormap, which focuses on the
oscillatory threshold, reveals an interesting phenomenon. The threshold
related to positive excitation currents (solid bold line) is shifted compared
to the one related to negative excitation currents (dashed bold line). This
mismatch between the thresholds creates hybrid oscillatory zones, in which the
DC-SQUID\ is driven to the oscillatory zone either for positive currents or
negative currents, but not for both.

Figure \ref{EMb42aExp}$\mathrm{(b)}$ plots a colormap of the voltage noise
level, measured using the spectrum analyzer at a frequency of $1%
\operatorname{kHz}%
$ (any frequency which is not an integer harmonics of $\omega_{\mathrm{x}}$
gives similar results). Plotting the noise level is a good way to detect
thresholds since noise-rise is generally expected near any bifurcation
threshold \cite{Huberman1980_750}. Two distinct patterns of noise-rise are
clearly seen in the colormap. One is related to the solid line positive
threshold and the other to the dashed line negative threshold. Note that this
noise-rise is almost undetectable in the lock-in measurements.

\begin{figure}
[ptb]
\begin{center}
\includegraphics[
height=2.7048in,
width=3.3441in
]%
{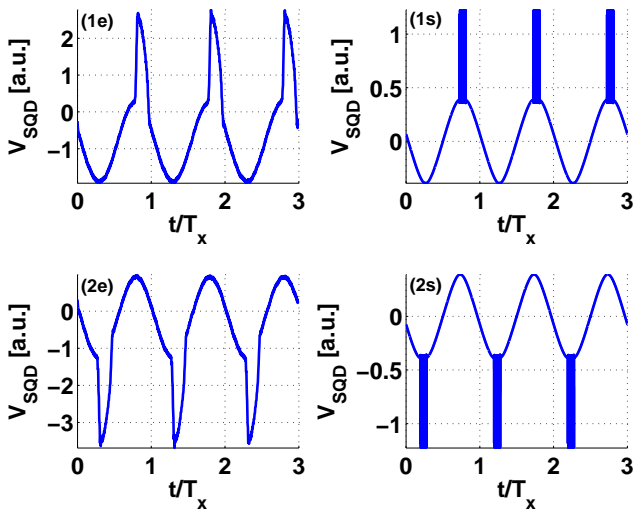}%
\caption{Time domain experimental measurements (left column) and numerical
simulation (right column) for E42. The two rows correspond to the two sets of
control parameters marked in Fig. \ref{EMb42aExp}.}%
\label{E42a_SimExp_TD}%
\end{center}
\end{figure}

The existence of hybrid zones is easily observed in time domain measurements.
Figure \ref{E42a_SimExp_TD} shows two pairs of time domain traces, each pair
has an experimental trace (panels $(\mathrm{1e})$, $(\mathrm{2e})$) and a
simulated trace (panels $(\mathrm{1s})$, $(\mathrm{2s})$), measured
(calculated) for the parameters labeled by plus marks and the corresponding
number in Fig. \ref{EMb42aExp}$\mathrm{(a)}$ and \ref{EMb42aExp}$\mathrm{(b)}%
$. The first (second) pair is related to the hybrid zone where only positive
(negative) currents drive the system to the oscillatory zone. Accordingly, the
first (second) pair experiences only positive (negative) spikes. The
difference between the measured and simulated line shapes of the spikes might
be due to finite (about $%
\operatorname{MHz}%
$) bandwidth of our measurement setup which is far too low to resolve high
frequency spikes. Therefore, measured spikes in the output signal merge into
one continuous and slowly decaying pulse. Local heating effects, which are
neglected in this simulation and will be discussed later, may also degrade
DC-SQUID\ performance and suppress the spikes.%

A statistical analysis of the time-domain behavior is seen in Fig.
\ref{EMb42aExp}$\mathrm{(c)}$, which plots the probability of counting a
positive spike $P_{\mathrm{s}}^{\mathrm{pos}}$, minus the probability of
counting a negative one $P_{\mathrm{s}}^{\mathrm{neg}}$, during a single
lock-in excitation cycle. The probabilities were measured for the parameter
range spanned by $I_{\mathrm{x}}$ and $\Phi_{\mathrm{x}}$, and were calculated
on $2\sec$ long traces with $\omega_{\mathrm{x}}/2\pi=2.5%
\operatorname{kHz}%
$. This colormap clearly reveals the existence of the hybrid oscillatory
zones, where the differential probability of counting spike having either
positive or negative polarities is almost one. The direction of the spikes
agrees with the prediction of the stability diagram. Outside the hybrid zones
the colormap shows near zero differential probability. In these areas there
are no spikes at all, or the counting of positive and negative spikes is
similar. This behavior could be used for creating bidirectional DC-SQUID
sensors in which the polarity of measured voltage indicates the polarity of
the detected flux change. Panel $\mathrm{(d)}$ plots the second harmonics of
the measured SQUID\ voltage (at $5%
\operatorname{kHz}%
$). This harmonics is expected to be amplified when the measured voltage is
asymmetric in time, i.e. neither symmetric nor anti-symmetric time response.
The hybrid zones are characterized by such a time response, and indeed, the
plotted colormap shows strong response of the second harmonics in those zones.

\section{Parametric Excitation}

In recent years several demonstrations of using SQUIDs to manipulate the
resonance frequencies of a superconducting resonator were reported
\cite{Suchoi2010_174525,Sandberg2008a,Castellanos-Beltran2007_083509,Laloy2008}%
. A SQUID in such applications is usually considered as a nonlinear variable
inductor embedded in the resonator in a way that couples the resonance
frequencies of the resonator to the SQUID impedance. The variation of the
SQUID inductance is usually done by changing the magnetic flux through the
SQUID, whereas the current through the SQUID is defined by the state of the
coupled system.

\subsection{Stability Zones}

In the following, we analyze the stability of a DC-SQUID, excited by a
magnetic flux having constant and alternating parts. Consider the case where
$\Phi_{\mathrm{x}}$ is given by%
\begin{equation}
\Phi_{\mathrm{x}}=\Phi_{0}\left(  \Phi_{\mathrm{x}}^{\mathrm{dc}}%
+\Phi_{\mathrm{x}}^{\mathrm{ac}}\cos\left(  \omega_{\mathrm{px}}t\right)
\right)  , \label{xParamFlux}%
\end{equation}
where $\Phi_{\mathrm{x}}^{\mathrm{dc}}/\Phi_{0}$ and $\Phi_{\mathrm{x}%
}^{\mathrm{ac}}/\Phi_{0}$ are arbitrary amplitudes. Recall that for the case
of $\beta_{\mathrm{L}}\gg1$, and for the minimum point near $\left(
\gamma_{+},\gamma_{-}\right)  =\left(  0,0\right)  $, the bounding rectangle
crosses the axis $I_{\mathrm{x}}=0$ at the points $\Phi_{\mathrm{x}}=\pm
\Phi_{0}\widetilde{\beta}_{\mathrm{L}}/2\pi$, where $\widetilde{\beta
}_{\mathrm{L}}=\left(  1-\alpha\right)  \beta_{\mathrm{L}}$. The range of
stability for the minima points near $\left(  \gamma_{+},\gamma_{-}\right)
=\left(  n\pi,n\pi\right)  $, where $n$ is integer, is given by%
\begin{equation}
-\frac{\widetilde{\beta}_{\mathrm{L}}}{2\pi}+n\leq\frac{\Phi_{\mathrm{x}}%
}{\Phi_{0}}\leq\frac{\widetilde{\beta}_{\mathrm{L}}}{2\pi}+n.
\label{paramStabContours}%
\end{equation}

Furthermore, the stability condition is achieved when the largest value of
$\Phi_{\mathrm{x}}$, i.e. $\Phi_{\mathrm{x}}^{\mathrm{dc}}+\Phi_{\mathrm{x}%
}^{\mathrm{ac}}$, coincides with the largest value of the stability range,
i.e. $\widetilde{\beta}_{\mathrm{L}}/2\pi$, or when the smallest value of
$\Phi_{\mathrm{x}}$, i.e. $\Phi_{\mathrm{x}}^{\mathrm{dc}}-\Phi_{\mathrm{x}%
}^{\mathrm{ac}}$, coincides with the smallest value of the stability range,
i.e. $-\widetilde{\beta}_{\mathrm{L}}/2\pi$. Thus the boundary contours in the
plan of $\Phi_{\mathrm{x}}^{\mathrm{dc}}$ and $\Phi_{\mathrm{x}}^{\mathrm{ac}%
}$ are given by the two equations%

\begin{equation}
\Phi_{\mathrm{x}}^{\mathrm{dc}}\pm\Phi_{\mathrm{x}}^{\mathrm{ac}}=\pm\Phi
_{0}\frac{\widetilde{\beta}_{\mathrm{L}}}{2\pi}+n\Phi_{0}. \label{PAstabCon2}%
\end{equation}

In experiments with resonators we employ the parametric amplification method
of operation
\cite{CASTELLANOS-BELTRAN2008,Castellanos-Beltran2009_944,Tholen2009_014019,Yamamoto2008_042510}%
. We inject a relatively weak probing signal into the resonator, having
frequency equals to one of the resonance frequencies of the resonator,
$\omega_{\mathrm{x}}$, and measure the reflected power off the resonator using
a spectrum analyzer. Note that the DC connections do not play any role in this
measurement. A weak signal is a one for which the current generated through
the SQUID is much smaller than its critical current. In addition, we applied
constant and variable magnetic flux through the SQUID, given by Eq.
(\ref{xParamFlux}) with $\omega_{\mathrm{px}}=\left(  2\omega_{\mathrm{x}%
}+\Delta\omega\right)  $, where $\Delta\omega$ is taken to be much smaller
than the resonance bandwidth of the resonator. The measured reflected power
spectrum includes a tone at $\omega_{\mathrm{x}}$ and several sidebands spaced
by $\pm m\cdot\Delta\omega$ from $\omega_{\mathrm{x}}$, where $m$ is an
integer. These sidebands are the products of the nonlinear frequency mixing
between $\omega_{\mathrm{x}}$ and $\omega_{\mathrm{px}}$. Although this mixing
process is more complex than our direct SQUID measurements, it should
essentially follow the same SQUID dynamics, provided that adiabatic
approximation is not violated, namely $\omega_{\mathrm{x}}\ll\omega
_{\mathrm{pl}}$. Therefore, we expect the various tones to reflect that dynamics.

Figure \ref{paramStabDiag}$\mathrm{(a)}$ shows the folded stability diagram,
drawn in the plane of $\Phi_{\mathrm{x}}^{\mathrm{dc}}$ and $\Phi_{\mathrm{x}%
}^{\mathrm{ac}}$ using Eq. (\ref{PAstabCon2}) and $\widetilde{\beta
}_{\mathrm{L}}=45$. The solid and dashed lines represent stability thresholds
for positive and negative polarities of $\Phi_{\mathrm{x}}^{\mathrm{ac}}$,
respectively. Four pairs of solid-dashed lines are drawn, each corresponds to
a different number of flux quanta trapped in the SQUID. The black bold line
marks the threshold between the PNDSZ and the PDSZ. Namely, for excitation
amplitudes $\left\vert \Phi_{\mathrm{x}}^{\mathrm{ac}}\right\vert $ smaller
than the black line, the SQUID will reach a LSZ after a finite number of
excitation periods. On the other hand for $\left\vert \Phi_{\mathrm{x}%
}^{\mathrm{ac}}\right\vert $ higher than the black line the SQUID will
periodically jump between LSZs. Note that in this method of excitation, where
the injected current is kept much smaller than the critical current, the SQUID
would not be driven to the oscillatory zone, even by an arbitrarily large
$\Phi_{\mathrm{x}}^{\mathrm{ac}}$.%
\begin{figure}
[ptb]
\begin{center}
\includegraphics[
height=2.4483in,
width=3.3715in
]%
{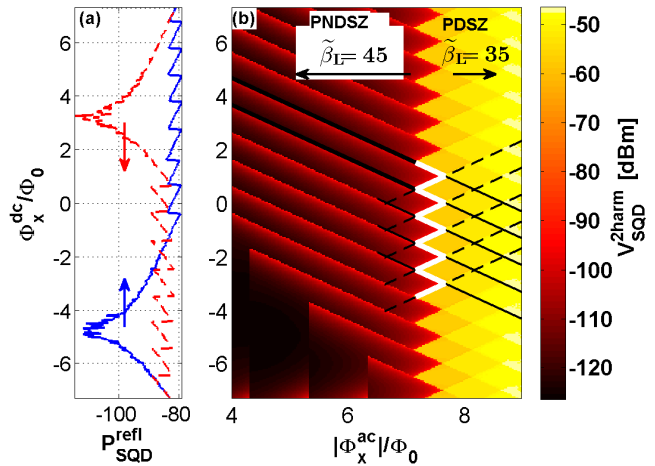}%
\caption{(Color online) Parametric excitation experimental $\mathrm{(a)}$ and
simulation $\mathrm{(b)}$ results. $\mathrm{(a)}$ Reflected power of the
second voltage harmonics as a function of increasing (blue) and decreasing
(red) DC flux. $\mathrm{(b)}$ Simulated second voltage harmonics in the plane
of $\Phi_{\mathrm{x}}^{\mathrm{ac}}$ and $\Phi_{\mathrm{x}}^{\mathrm{dc}}$.
The black contours mark the corresponding stability diagram drawn for
$\widetilde{\beta}_{\mathrm{L}}=45$. White bold contours mark the threshold to
the PDSZ.}%
\label{paramamp_emb42_sim_noheat}%
\end{center}
\end{figure}

Figure \ref{paramamp_emb42_sim_noheat}$\mathrm{(b)}$ shows the simulated
second voltage harmonics across the SQUID, i.e. at frequency $\omega
=2\omega_{\mathrm{px}}$, and for $\Delta\omega=0$, obtained using E42
parameters. The black contours, which are similar to the ones in Fig.
\ref{paramStabDiag}$\mathrm{(a)}$, mark the stability diagram. The distinction
between the periodic non-dissipative and periodic dissipative static zones is
clear, and is marked by the white bold PDSZ threshold. Like before with the
SQUID\ of E38, the PDSZ is characterized by relatively high SQUID\ voltage,
which is divided into diamond shaped regions. The PNDSZ, unlike with E38, is
also characterized by strong response along the stability contours. When one
such contour is crossed, a single transition between LSZ occurs. After this
transition no additional transitions would periodically occur. Nevertheless,
these boundaries are detectable due to the fact that the inductance of the
SQUID before and after a transition is different, and also due to the fact
that the excitation frequency is high, thus the effect of changes in the
SQUID\ inductance is measurable. Only transitions across the solid stability
contours are observed in the colormap of Fig. \ref{paramamp_emb42_sim_noheat}.
The reason is the sequence of the measurement, which includes sweeping the DC
flux monotonically up and down in the inner simulation loop. The colormap is
obtained while the flux amplitude is increased, thus transitions over the
solid contours are recorded; whereas the decreasing section is only used to
maintain the consecutiveness of initial condition, similar to the experiments.

Figure \ref{paramamp_emb42_sim_noheat}$\mathrm{(a)}$ shows measurement results
obtained from E42. A probe tone having frequency of $\omega_{\mathrm{x}}$ is
injected into the resonator, and the reflected power of the second order
sideband, i.e. the tone at $\omega_{\mathrm{x}}+\Delta\omega$, is measured,
while the DC\ flux is swept up (blue curve, marked by up headed arrow) and
down (red curve, marked by down headed arrow), and while keeping the AC flux
at a fixed amplitude. Following the blue curve from bottom to top, the
reflected power experiences a resonance-like absorption followed by a saw
teeth pattern. The resonance absorption pattern originates from the tuning of
the resonance frequency of the resonator relatively to the frequency of the
probing signal, thus effectively sweeping the probe in and out of resonance.
This sweeping is caused by the SQUID, which has flux dependent inductance
\cite{Suchoi2010_174525}, even when stuck in a single LSZ. After the SQUID is
driven across a stability threshold it falls to a new LSZ, into a location
which is one flux quantum away from the corresponding stability contour. Thus,
if the external flux is further increased, it would further drives the SQUID
to the same direction, and additional transitions would occur, spaced apart by
one flux quantum. If, on the other hand, the direction of the sweep changes
(red curve), the SQUID would first have to be driven across a whole LSZ, until
reaching the opposite stability threshold. Therefore, no matter where the flux
sweep changes direction along the saw teeth pattern, the reflected power would
experiences a resonance-like absorption followed by saw teeth pattern.

Looking back at Fig. \ref{paramamp_emb42_sim_noheat}$\mathrm{(b)}$, the
sweeping of $\Phi_{\mathrm{x}}^{\mathrm{dc}}$ up and down usually drives back
the SQUID to its original LSZ. When the AC flux amplitude $\Phi_{\mathrm{x}%
}^{\mathrm{ac}}$ is increased, the borders of a LSZ enclose one another. As a
result, the sweep across the first stability zone becomes shorter, and hence
the resonance absorption pattern becomes narrower along the DC flux axis. This
narrowing adds saw teeth along the sweep range, and in addition, the
SQUID\ might no longer return to its original LSZ, but rather to a new LSZ,
which correspond to a change of one in the number of trapped flux quanta. This
creates the sharp transition observed in the resonance-like absorption
patterns along the horizontal axis.

\subsection{Temperature dependant critical current}

When a SQUID is embedded in a resonator, the current $I_{\mathrm{x}}$ flowing
through the SQUID is driven by the resonator, and its frequency, thus equals
to one of the resonance frequencies of the resonator. These first few
frequencies are only about two to three orders of magnitude lower than the
plasma frequency of the SQUID \cite{Suchoi2010_174525}. Furthermore, the heat
transfer rate corresponding to hot-spots in the NBJJs may be of the order as
those frequencies or even slower \cite{Tarkhov2008_241112}, and therefore
effect of heating on the dynamics of the SQUID becomes significant. Assuming
for simplicity that the temperature $T_{k}$ ($k=1,2$) in each junction is
uniform. The dependence of the critical current on the temperature is given by
\cite{Skocpol1976_1045}%
\begin{equation}
\frac{I_{\mathrm{c}k}}{I_{\mathrm{c}0k}}=y\left(  \Theta_{k}\right)
\equiv\frac{\widetilde{y}\left(  \Theta_{k}\right)  }{\widetilde{y}\left(
\Theta_{0}\right)  },\label{IcVsT}%
\end{equation}
where $I_{\mathrm{c}0k}$ is the critical current of $k^{\mathrm{th}}$ NBJJ at
base temperature $T_{0}$ of the coolant, $\Theta_{k}=T_{k}/T_{\mathrm{c}}$ is
normalized temperature of the NBJJ (with respect to its critical temperature),
$\Theta_{0}=T_{0}/T_{\mathrm{c}}\,$, and where the function $\widetilde{y}$ is
given by%
\begin{equation}
\widetilde{y}\left(  \Theta\right)  =\left(  1-\Theta^{2}\right)
^{3/2}\left(  1+\Theta^{2}\right)  ^{1/2}.
\end{equation}
The $k^{\mathrm{th}}$ ($k=1,2$) NBJJ heat balance equations read%
\begin{equation}
C_{k}\frac{\mathrm{d}T_{k}}{\mathrm{d}t}=Q_{k}-H_{k}\left(  T_{k}%
-T_{0}\right)  ,\label{dT/dt}%
\end{equation}
where $C_{k}$ is thermal heat capacity, $Q_{k}=V_{k}^{2}/R_{\mathrm{J}}$ is
heating power, and $H_{k}$ is heat transfer coefficient. By using the notation
$\beta_{\mathrm{C}k}=2\pi C_{k}T_{\mathrm{c}}/\Phi_{0}I_{\mathrm{c}0}$ and
$\beta_{\mathrm{H}k}=H_{k}C_{k}\omega_{\mathrm{p}}$, Eq. \ref{dT/dt} becomes
\
\begin{equation}
\dot{\Theta}_{k}=\frac{\beta_{\mathrm{D}}}{\beta_{\mathrm{C}k}}\dot{\gamma
}_{k}^{2}-\beta_{\mathrm{H}k}\left(  \Theta_{k}-\Theta_{0}\right)
.\label{EOM_Theta1}%
\end{equation}
%

\begin{figure}
[ptb]
\begin{center}
\includegraphics[
height=2.4848in,
width=3.3723in
]%
{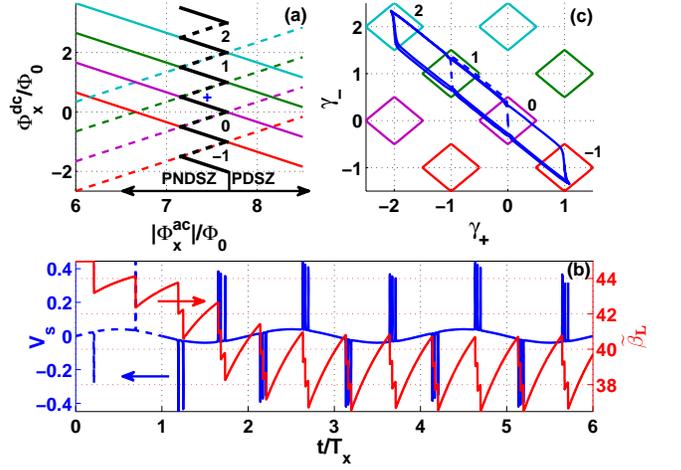}%
\caption{(Color online) $\mathrm{(a)}$ Stability zones in the plan of the
control parameters $\Phi_{\mathrm{x}}^{\mathrm{ac}}$ and $\Phi_{\mathrm{x}%
}^{\mathrm{dc}}$, drawn for $\widetilde{\beta}_{L}=45$. The curves are drawn
using solid lines for $\Phi_{\mathrm{x}}^{\mathrm{ac}}>0$ and dashed lines for
$\Phi_{\mathrm{x}}^{\mathrm{ac}}<0$. The numbers beside each pair crossing
represent the total magnetic flux quanta trapped in the SQUID in the
corresponding LSZ (in units of $\Phi_{0}$). The bold black curve marks the
PDSZ threshold. Simulation results of the SQUID voltage and $\widetilde{\beta
}_{\mathrm{L}}$ in the normalized time domain $\mathrm{(b)}$, and $\gamma_{+}$
and $\gamma_{-}$ in the phase space $\mathrm{(c)}$. The simulation is
calculated using the set of control parameters marked by the plus sign in
panel $\mathrm{(a)}$. Time is normalized by the period of probing tone
injected to the resonator, $T_{\mathrm{x}}$, }%
\label{paramStabDiag}%
\end{center}
\end{figure}
Figure \ref{paramStabDiag}, panels $\mathrm{(b)}$ and $\mathrm{(c),}$ show
simulation results of the SQUID voltage and $\widetilde{\beta}_{\mathrm{L}}$
in the time domain $\mathrm{(b),}$ and $\gamma_{+}$ and $\gamma_{-}$ in the
phase space $\mathrm{(c)}$, calculated for the control parameters marked by
the plus sign in Fig. \ref{paramStabDiag}$\mathrm{(a)}$. The temperature in
this simulation is not held at the base temperature but rather evolves
according to Eq. (\ref{EOM_Theta1}). The parameters $\beta_{\mathrm{C}k}=$
$320$ and $\beta_{\mathrm{H}k}=$ $5\times10^{-4}$ ($k=1,2$) were calculated
analytically according to Refs.
\cite{kinInd_Johnson96,HED_Weiser81,Monticone1999_3866}. Further explanations
about parameters calculation is found in appendix $\mathrm{B}$ of Ref.
\cite{Suchoi2010_174525}.\ One expects that for the chosen control parameters
the SQUID would oscillates between two LSZs. The dynamics plotted in panel
$\mathrm{(c)}$, however, shows that this is true only during the first
excitation cycle (dashed line), and that afterwards the SQUID oscillates
between four LSZs. Such a change in the dynamic behavior of the SQUID is also
observed in the time domain voltage trace plotted in panel $\mathrm{(b)}$. The
number of voltage spikes increases from one during the first excitation cycle
(dashed line) to three in the end of the second excitation cycle (solid line).
This behavior can be explained by the dynamical change in the value of
$\widetilde{\beta}_{\mathrm{L}}$, plotted as red curve in panel $\mathrm{(b)}%
$. The hysteretic parameter $\widetilde{\beta}_{\mathrm{L}}$ experiences
relaxation oscillations having their mean value changing during the first
three excitation cycles. The relaxation oscillations are driven by the voltage
spikes which dissipate energy and produce heat. This heat increases the
temperature of the NBJJs, which in turn decreases their critical current
according to Eq. (\ref{IcVsT}), thus decreasing the value of $\beta
_{\mathrm{L}}$. The reduction in $\beta_{\mathrm{L}}$ results in a shift of
the stability diagram towards the origin, and consequently the given set of
control parameters effectively drives the SQUID between increased numbers of LSZs.%

\begin{figure}
[ptb]
\begin{center}
\includegraphics[
height=2.3562in,
width=3.3441in
]%
{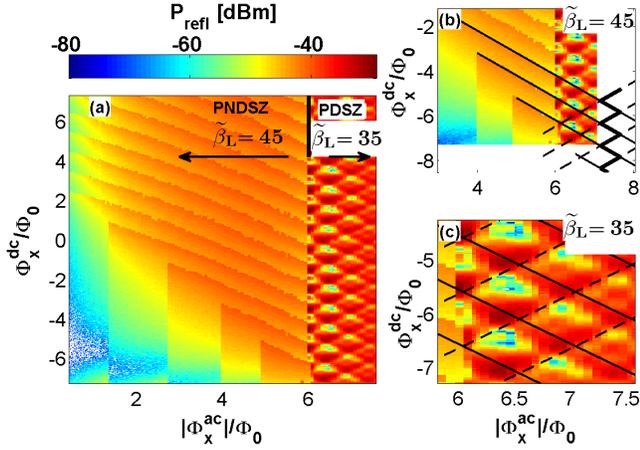}%
\caption{(Color online) Parametric excitation experimental results. The
colormaps show the reflected power off the resonator in the plane of
$\Phi_{\mathrm{x}}^{\mathrm{ac}}$ and $\Phi_{\mathrm{x}}^{\mathrm{dc}}$.
Panels $\mathrm{(b)}$ and $\mathrm{(c)}$ show the fitting of the stability
diagram to the measured data using $\widetilde{\beta}_{\mathrm{L}}=45$ and
$\widetilde{\beta}_{\mathrm{L}}=35$, respectively.}%
\label{paramExc_emb42}%
\end{center}
\end{figure}

Figure \ref{paramExc_emb42} shows the reflected power of the second order
mixing product, i.e. the tone at $\omega_{\mathrm{x}}+\Delta\omega$, measured
in the plane of $\Phi_{\mathrm{x}}^{\mathrm{dc}}$ and $\Phi_{\mathrm{x}%
}^{\mathrm{ac}}$ from E42. This measurement corresponds to the simulation
shown in Fig. \ref{paramamp_emb42_sim_noheat}. The figure contains three
panels, where panels $\mathrm{(b)}$ and $\mathrm{(c)}$ show partial sections
of the main colormap shown in panel \textrm{(a)}, each with a corresponding
fit of the stability diagram. Looking at panel $\mathrm{(a)}$, most of the
observations agree well with the simulated results, seen in Fig.
\ref{paramamp_emb42_sim_noheat}. The PDSZ is characterized by strong
reflection response and stability regions having diamond shapes. The PNDSZ
experiences strong response along the boundaries of the LSZs. The response
lines slightly bend for low excitation amplitudes ($\Phi_{\mathrm{x}%
}^{\mathrm{ac}}\lesssim3\Phi_{0}$), but follow rather strait line for higher
amplitudes. Good fitting of the stability diagram to these lines is achieved
with $\widetilde{\beta}_{\mathrm{L}}=45$, as shown in panel $\mathrm{(b)}$.
The bending of the boundary lines at low amplitudes might be due to slow rise
of the average temperatures of the SQUID. The reason why only transitions
across solid stability contours are measured is the measurement protocol,
which includes a sweep of the DC flux up and down in the inner measurement
loop, while measurements are recorded only during the incremental duration of
the sweep. In addition to the emerging of the LSZ boundaries in the
measurement, also the influence of the variation of the SQUID inductance
within a LSZ on the reflected power is observed. Furthermore, the sharp
transitions along the horizontal axis, corresponding to DC flux sweeps that do
not drive the SQUID back to its original LSZ, are observed at $\Phi
_{\mathrm{x}}^{\mathrm{ac}}/\Phi_{0}=[1.4,2.75,4.05,4.74]$, in agreement with
the simulation results.

The stability diagram in Fig. \ref{paramExc_emb42}$\mathrm{(b)}$, drawn for
$\widetilde{\beta}_{\mathrm{L}}=45$, predicts that the PDSZ threshold should
zigzag along the stability contours in the range of $\Phi_{\mathrm{x}%
}^{\mathrm{ac}}/\Phi_{0}\in\lbrack7.5,8]$. The measured colormap, however,
shows that the PDSZ threshold passes along a strait vertical line starting at
point $(\Phi_{\mathrm{x}}^{\mathrm{dc}}=-6.25\Phi_{0},\Phi_{\mathrm{x}%
}^{\mathrm{ac}}=5.95\Phi_{0})$. A fitting process of the stability diagram to
the PDSZ section itself produces $\widetilde{\beta}_{\mathrm{L}}\mathbf{=}35$
(see Fig. \ref{paramExc_emb42}$\mathrm{(c))}$. This duality in the value of
$\beta_{\mathrm{L}}$ can be understood if changes of the SQUID temperature are
taken into account. The PDSZ threshold point exactly coincides with one of the
LSZ threshold contours. The heat generated in single transition across that
threshold momentarily decreases $\widetilde{\beta}_{\mathrm{L}}$. An
additional transition may be triggered provided that the relaxation of the
first one lasts long enough, which in turn may cause further heating of the
DC-SQUID. Eventually a new mean temperature is achieved for which
$\widetilde{\beta}_{\mathrm{L}}=35$. The measurement protocol dictates that
this new temperature would be kept and that the DC-SQUID would stay in the
PDSZ for the rest of the measurement. Note that the initial value of
$\widetilde{\beta}_{\mathrm{L}}=45$ differs from the value of $\widetilde
{\beta}_{\mathrm{L}}=80$ that was measured\ using lock-in amplifier. This can
be explained by the local heating that the high frequency flux excitation
induces in the DC-SQUID, especially in the NBJJs, through the dissipation of
circulating current due to RF surface resistance. Such reduction in
$\widetilde{\beta}_{\mathrm{L}}$ from $80$ to $45$ may be induced by a change
of the local temperature by approximately $2%
\operatorname{K}%
$.

\subsection{The Case of RF SQUID Parametric Excitation}

The stability diagram for a DC-SQUID can be evaluated numerically, or be
analytically approximated for the extreme cases of $\beta_{\mathrm{L}}\ll1$
and $\beta_{\mathrm{L}}\gg1$. For RF-SQUID, on the other hand, it could be
exactly evaluated analytically. Consider a RF-SQUID having self-inductance
$L$, critical current $I_{\mathrm{c}}$, and externally applied magnetic flux
$\Phi_{\mathrm{x}}=\left(  \Phi_{0}/2\pi\right)  \phi_{\mathrm{x}}$. The
dynamics of the total magnetic flux $\Phi=\left(  \Phi_{0}/2\pi\right)  \phi$
threading the RF-SQUID loop is governed by the potential energy $U=U_{0}%
u_{\mathrm{RF}}$ \cite{NAMR_Buks06}, where
\begin{equation}
u_{\mathrm{RF}}=\left(  \phi-\phi_{\mathrm{x}}\right)  ^{2}-2\beta
_{\mathrm{L}}\cos\phi,
\end{equation}
and $U_{0}=\Phi_{0}^{2}/\left(  8\pi^{2}L\right)  $. A local minimum point of
$u_{\mathrm{RF}}$ is found by solving $\partial u_{\mathrm{RF}}/$
$\partial\phi=0$ and requiring that $\partial^{2}u/\partial\phi^{2}>0$.
Clearly, if $\phi_{\mathrm{m}}$ is a local minimum point of the potential
$u_{\mathrm{RF}}$ with a given $\phi_{\mathrm{x}}$, then $\phi_{\mathrm{m}%
}+2n\pi$ is also a local minimum point of the potential $u_{\mathrm{RF}}$ with
an externally applied flux of $\phi_{\mathrm{x}}+2n\pi$, provided that $n$ is integer.

Lose of stability occurs when $\partial^{2}u/\partial\phi^{2}=0$, namely when
$\cos\phi=-1/\beta_{\mathrm{L}}$. This can occur only when $\beta_{\mathrm{L}%
}>1$, since otherwise the system is expected to be monostable. This condition
is satisfied when $\phi_{\mathrm{x}}=\phi_{\mathrm{x,b}}$, where%
\begin{equation}
\phi_{\mathrm{x,b}}=\pi-\arccos\left(  \frac{1}{\beta_{\mathrm{L}}}\right)
+\sqrt{\beta_{\mathrm{L}}^{2}-1}.
\end{equation}

For a general integer $n$, the local minimum of the potential $u_{\mathrm{RF}%
}$ near $\phi=2n\pi$ remains stable in the range%
\begin{equation}
-\phi_{\mathrm{x,b}}+2n\pi\leq\phi_{\mathrm{x}}\leq\phi_{\mathrm{x,b}}+2n\pi.
\end{equation}

Similarly to the case of DC-SQUID, where $\Phi_{\mathrm{x}}$ is given by Eq.
(\ref{xParamFlux}), the boundary contours in the plan of $\Phi_{\mathrm{x}%
}^{\mathrm{dc}}$ and $\Phi_{\mathrm{x}}^{\mathrm{ac}}$ are given by the two
equations
\begin{equation}
\Phi_{\mathrm{x}}^{\mathrm{dc}}\pm\Phi_{\mathrm{x}}^{\mathrm{ac}}=\Phi
_{0}\left(  \pm\frac{\phi_{\mathrm{x,b}}}{2\pi}+n\right)
,\label{RF_stab_Con_p}%
\end{equation}
for the largest and smallest values of $\Phi_{\mathrm{x}}$, respectively.%

\begin{figure}
[ptb]
\begin{center}
\includegraphics[
height=2.5737in,
width=3.3441in
]%
{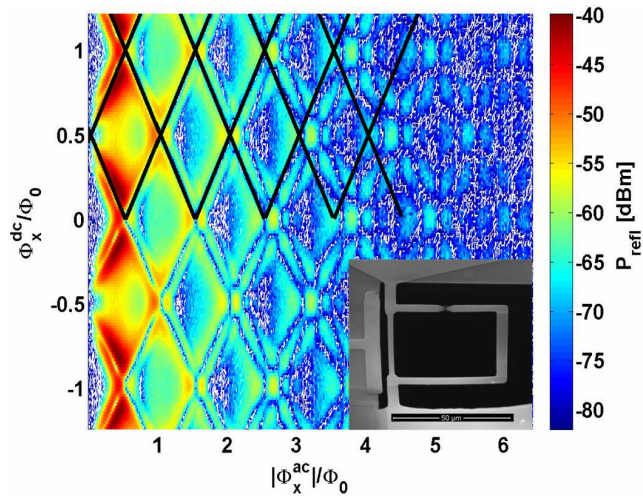}%
\caption{(Color online) Parametric excitation experimental results using E19,
which has an integrated RF-SQUID. The colormap shows the reflected power off
the resonator in the plane of $\Phi_{\mathrm{x}}^{\mathrm{ac}}$ and
$\Phi_{\mathrm{x}}^{\mathrm{dc}}$. The black contours mark the corresponding
stability diagram. They are applied only on half of the colormap in order to
leave some data uncovered. The inset shows a SEM image of a RF SQUID.}%
\label{paramAmpRFsqd}%
\end{center}
\end{figure}

Figure \ref{paramAmpRFsqd} shows the reflected power of the third order mixing
product, i.e. the tone at $\omega_{\mathrm{x}}+2\Delta\omega$, measured in the
plane of $\Phi_{\mathrm{x}}^{\mathrm{dc}}$ and $\Phi_{\mathrm{x}}%
^{\mathrm{ac}}$ with E19, which has an embedded RF-SQUID instead of a
DC-SQUID. The black contours represent the stability diagram, plotted using
Eqs. (\ref{RF_stab_Con_p}) for $\beta_{\mathrm{L}}=1.5$. The stability diagram
qualitatively matches the measured data for the first few stability diamonds.

Note that the value of $\beta_{\mathrm{L}}$ for the RF-SQUID\ of E19 is more
than order of magnitude smaller than that of E38 and E42. This might be due to
the fact that this RF-SQUID is fully fabricated on Silicon Nitride membrane,
and thus has reduced thermal coupling to the coolant compared to the DC-SQUIDs
of E38 and \ E42. This in turn might lead to an increased local temperature of
the NBJJ, and to a degraded value of $\beta_{\mathrm{L}}$.

Note also that the diamonds themselves have distinct reflection patterns
inside their bounded zone (see Figs. \ref{paramExc_emb42} and
\ref{paramAmpRFsqd}), which are not perfectly periodic with $\Phi_{\mathrm{x}%
}^{\mathrm{ac}}$, but are reproducible in measurements. They are only detected
in measurements with resonators having either DC or RF SQUIDs, but not in
measurements done directly with DC-SQUIDs. Our model only handles the DC-SQUID
equations of motion, and thus cannot provide full description of the dynamics
of the resonator-SQUID system. Further theoretical work is needed for modeling
combined TLR-SQUID system, in order to fully understand these experimental results.

\section{Conclusions}

In conclusion, we have studied the response of a nano-bridge based SQUID
embedded in a superconducting microwave resonator. nano-bridge based SQUIDs
are usually characterized by high critical current, and thus enhanced
metastable and hysteretic response. Several phenomena were observed, including
periodic dissipative static zone in which periodic transitions between local
stable state occur; hybrid oscillatory zones, in which the SQUID is driven to
the oscillatory zone by one polarity of the excitation amplitude but not for
the other, and dynamical variations in $\beta_{\mathrm{L}}$ due to the effect
of self-heating. The behaviors of the SQUIDs were compared with theory both
analytically and numerically with good agreement.

E.S. is supported by the Adams Fellowship Program of the Israel Academy of
Sciences and Humanities. This work is supported by the German Israel
Foundation under grant 1-2038.1114.07, the Israel Science Foundation under
grant 1380021, the Deborah Foundation, the Poznanski Foundation, Russell
Berrie nanotechnology institute, Israeli Ministry of Science, the European
STREP QNEMS project, and MAFAT.

\appendix

\section{Nano-bridge current-phase relation}

The CPR of a single short channel of transmission $\tau$ is given by
\cite{Beenakker1991_3056}%
\begin{equation}
I=\frac{e\Delta}{2\hbar}J\left(  \gamma\right)  \;,
\end{equation}
where%
\begin{equation}
J\left(  \gamma\right)  =\frac{\tau\sin\gamma}{\sqrt{1-\tau\sin^{2}%
\frac{\gamma}{2}}}\;.
\end{equation}
The NBJJs in our devices are not ideal one dimensional point contacts as Ref.
\cite{Beenakker1991_3056} assumes, however, we have found that the above
simple analytical result resembles the CPR which is obtained by solving the
Ginzburg-Landau equation in the limit of short bridge (in comparison with the
coherence length) \cite{Likharev1979_101}. Let $\gamma_{0}$ be the point at
which the factor $J\left(  \gamma\right)  $ has its largest value $J\left(
\gamma_{0}\right)  $, which is given by%
\begin{equation}
J\left(  \gamma_{0}\right)  =2\sqrt{2\left(  1-\sqrt{1-\tau}\right)  -\tau}\;.
\end{equation}
Using this result the current $I$ can be written in terms of $I_{\mathrm{c}}$
as $I/I_{\mathrm{c}}=F\left(  \gamma\right)  $, where%
\begin{equation}
F\left(  \gamma\right)  =\frac{\tau\sin\gamma}{2\sqrt{2\left(  1-\sqrt{1-\tau
}\right)  -\tau}\sqrt{1-\tau\sin^{2}\frac{\gamma}{2}}}\;.
\end{equation}
Replacing the $\sin\gamma_{\mathrm{k}}$ terms ($k=1,2$) in Eqs.
(\ref{eom gamma_1}) and (\ref{eom gamma_2}) by $F\left(  \gamma_{\mathrm{k}%
}\right)  $ leads to the following modified EOMs%
\begin{gather}
\ddot{\gamma}_{1}+\beta_{\mathrm{D}}\dot{\gamma}_{1}+\left(  1+\alpha
_{0}\right)  y\left(  \Theta_{1}\right)  F\left(  \gamma_{1}\right)
\nonumber\\
+\frac{1}{\beta_{\mathrm{L0}}}\left(  \gamma_{1}-\gamma_{2}+2\pi
\Phi_{\mathrm{x}}/\Phi_{0}\right)  =I_{\mathrm{x}}/I_{\mathrm{c}%
0}+g_{\mathrm{n}1}, \label{EOM_CPR1}%
\end{gather}
and%
\begin{gather}
\ddot{\gamma}_{2}+\beta_{\mathrm{D}}\dot{\gamma}_{2}+\left(  1-\alpha
_{0}\right)  y\left(  \Theta_{2}\right)  F\left(  \gamma_{2}\right)
\nonumber\\
-\frac{1}{\beta_{\mathrm{L0}}}\left(  \gamma_{1}-\gamma_{2}+2\pi
\Phi_{\mathrm{x}}/\Phi_{0}\right)  =I/I_{\mathrm{c}0}+g_{\mathrm{n}2}.
\label{EOM_CPR2}%
\end{gather}

\bibliographystyle{apsrev}
\bibliography{Bibilography}

\end{document}